\documentclass[9pt,article,twoside]{rmaa-rho-class/rmaa-rho}
\RMxAAtemplatetype{\RMxAA} 

\usepackage{graphicx}	
\usepackage{amsmath}	
\usepackage{threeparttable}
\vol{62}
\pages{1-8}
\thisyear{2026}
\doi{\href{https://www.astroscu.unam.mx/rmaa/RMxAA..XX-X}{https://www.astroscu.unam.mx/rmaa/RMxAA..XX-X}}

\title{On the stellar parameter dependence of the combined Fe I and Fe II chromospheric emission-line in the wings of the Ca II K line}

\author[1]{Faiber Rosas-Portilla \orcidlink{0000-0002-5084-0898}}
\author[1]{Luis Zapata \orcidlink{0000-0003-2343-7937}}
\author[2]{Klaus-Peter Schr\"oder}
\author[2]{Sandra González-Enríquez \orcidlink{0009-0007-4058-5313}}
\author[2]{Dennis Jack \orcidlink{0000-0003-3343-6743}}
\author[3]{Robert Stencel}

\affil[1]{Instituto de Radioastronomía y Astrofísica, Universidad Nacional Autónoma de México, Morelia MICH 58090, México}
\affil[2]{Departamento de Astronomía, Universidad de Guanajuato, Guanajuato GTO 36023, México}
\affil[3]{Chamberlin Observatory, University of Denver, 2930 E Warren Ave., Denver CO 80208, USA}

\leadauthor{F. Rosas-Portilla et al.}
\smalltitle{On the stellar parameter dependence of the combined Fe I and Fe II chromospheric emission-line in the wings of the Ca II K line}

\corres{Faiber Rosas-Portilla}
\email{f.rosas@irya.unam.mx}

\received{\today}
\accepted{\today}

\license{Texto de la licencia aquí}

\setbool{rho-abstract}{true} 
\setbool{rho-resumen}{true} 

\begin{abstract}
We present an analysis of the chromospheric emission of \ion{Fe}{i} + \ion{Fe}{ii} in the wing of the \ion{Ca}{ii} K line of 21 G and K giant stars. Stellar parameters and absolute chromospheric fluxes were obtained by comparison with \textsc{phoenix} models. The iron-blend emission shows a proportionality with the \ion{Ca}{ii} K flux for stars with similar stellar parameters and also for individual stars observed at different epochs, confirming that both diagnostics seems to respond to the same underlying chromospheric conditions. We found a dependence of the iron-blend emission on stellar parameters by fitting a power-law relation. The exponent of the effective temperature statistically matches with that found for \ion{Ca}{ii} K; which, together with a metallicity dependence, indicate that the iron-blend may act as an indirect tracer of chromospheric thermal conditions. The chromospheric emission flux ratio (\ion{Fe}{i} + \ion{Fe}{ii})/\ion{Ca}{ii} K exhibits a distinct slope change around $\log g \approx 2.5$ dex; which causes the gravity dependence to be different for stars with gravities lower and higher than 2.5 dex. This behavior suggests a transition between regimes where the two diagnostics probe similar physical conditions and where they begin to diverge in the sensitivity to the chromospheric extension. We discuss the implications of this behavior for the thermal structure of the lower chromosphere and for the formation heights of both diagnostics.
\end{abstract}

\keywords{stars: chromospheres -- stars: late-type -- stars: activity -- stars: emission-line -- techniques: spectroscopic}

\begin{resumen}
Presentamos un análisis de la emisión cromosférica de \ion{Fe}{i} + \ion{Fe}{ii} en el ala de la línea \ion{Ca}{ii} K de 21 estrellas gigantes G y K. Los parámetros estelares y los flujos cromosféricos absolutos se obtuvieron por comparación con modelos de \textsc{phoenix}. La emisión de la mezcla de hierro muestra una proporcionalidad con el flujo de \ion{Ca}{ii} K para estrellas con parámetros estelares similares y también para estrellas individuales observadas en diferentes épocas, confirmando que ambos diagnósticos parecen responder a las mismas condiciones cromosféricas subyacentes. Encontramos una dependencia de la emisión de la mezcla de hierro con los parámetros estelares mediante el ajuste de una relación de ley de potencia. El exponente de la temperatura efectiva coincide estadísticamente con el encontrado para \ion{Ca}{ii} K; el cual, en conjunto con una dependencia de la metalicidad, indican que la mezcla de hierro puede actuar como un trazador indirecto de las condiciones térmicas cromosféricas. La relación de flujo de emisión cromosférica (\ion{Fe}{i} + \ion{Fe}{ii})/\ion{Ca}{ii} K exhibe un cambio de pendiente distintivo alrededor de $\log g \approx 2.5$ dex; causando que la dependencia de la de gravedad sea distinto para estrellas con gravedades menores y mayores a 2.5 dex. Este comportamiento sugiere una transición entre regímenes donde los dos diagnósticos sondean condiciones físicas similares y donde comienzan a divergir en sensibilidad a la extensión cromosférica. Discutimos las implicaciones de este comportamiento para la estructura térmica de la cromosfera inferior y para las alturas de formación de ambos diagnósticos.
\end{resumen}

\begin{document}

\maketitle
\pagestyle{fancy}\thispagestyle{firststyle}


\section{Introduction}

The presence of emission lines in the wings of the \ion{Ca}{ii} H\&K lines ($\lambda = 3968.47$ \AA{} and $3933.66$ \AA{} in air) has been a subject of interest in stellar chromosphere research since their discovery in the solar spectrum by \cite{Jewell1898}, and has motivated further research into such wing emission in the following decades \citep{Evershed1927, Thackeray1935, Canfield1971, Lites1974, Cram1980, Watanabe1986, Schmidt2013, Harnes2025}. A very comprehensive study was carried out by \cite{Stencel1977} using the large and homogeneous library of coudé spectrograms obtained by O.C. Wilson at the Mount Wilson and Palomar Observatory. He aimed for the metallic emission lines in the wings of the \ion{Ca}{ii} H\&K lines in a sample of 50 F-, G-, K- and M-type stars, and analyzed how the width of these lines varies with stellar luminosity. \cite{Stencel1977} showed that the metallic wing emission lines appear to be strong only when the chromospheric emission of \ion{Ca}{ii} K itself was, and similar to the Wilson-Bappu effect \citep{Wilson1957}, this iron wing emission could also serve as a distance indicator for highly luminous stars cooler than spectral type F2.

\begin{figure}
    \includegraphics[width=0.99\columnwidth]{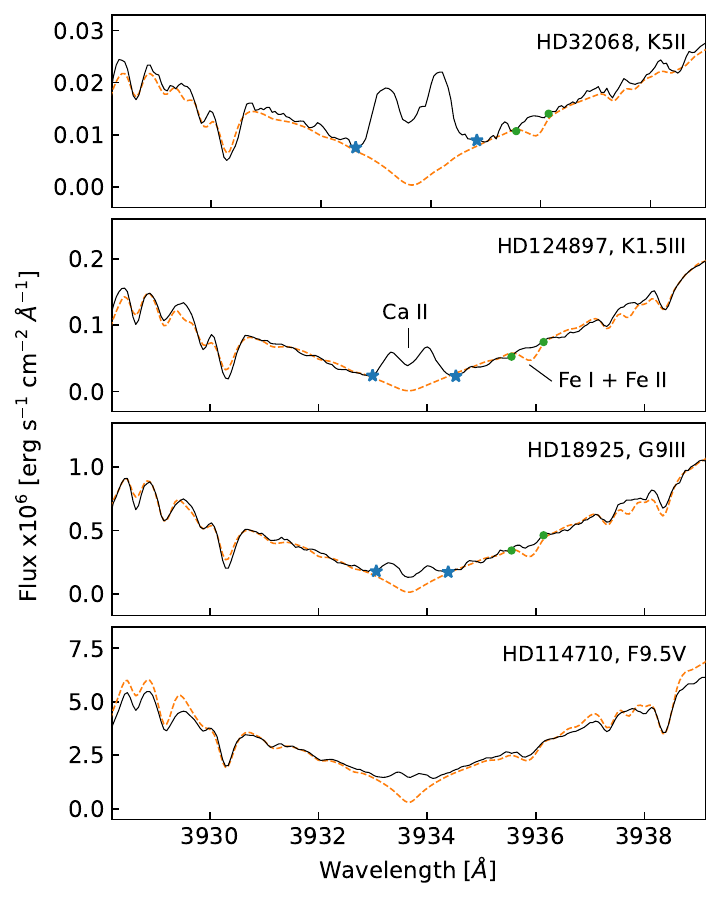}
    \caption{Combined iron emission lines of \ion{Fe}{i} + \ion{Fe}{ii} in the wings of \ion{Ca}{ii} K observed in TIGRE-HEROS spectra. The orange dashed line corresponds to the \textsc{phoenix} photospheric LTE model for each star. The points represent the wavelength range in which the fluxes of \ion{Ca}{ii} (blue asterisk) and \ion{Fe}{i} + \ion{Fe}{ii} (green points) were measured.}
    \label{fig:Synthetic_vs_Observed_Fe_and_CaII}
\end{figure}

\begin{figure*}
    \includegraphics[width=0.99\columnwidth]{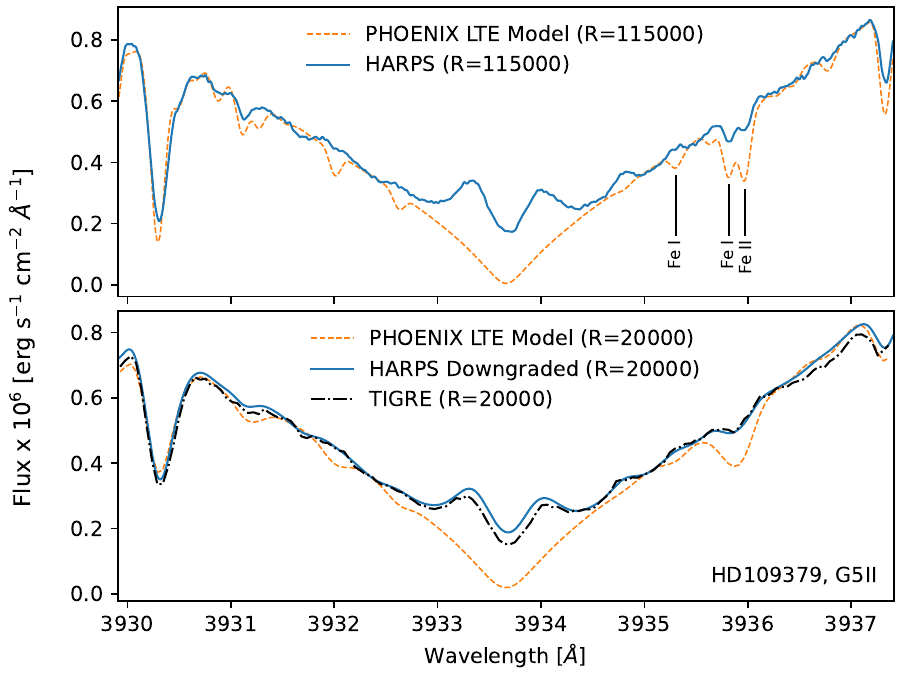}
    \includegraphics[width=0.99\columnwidth]{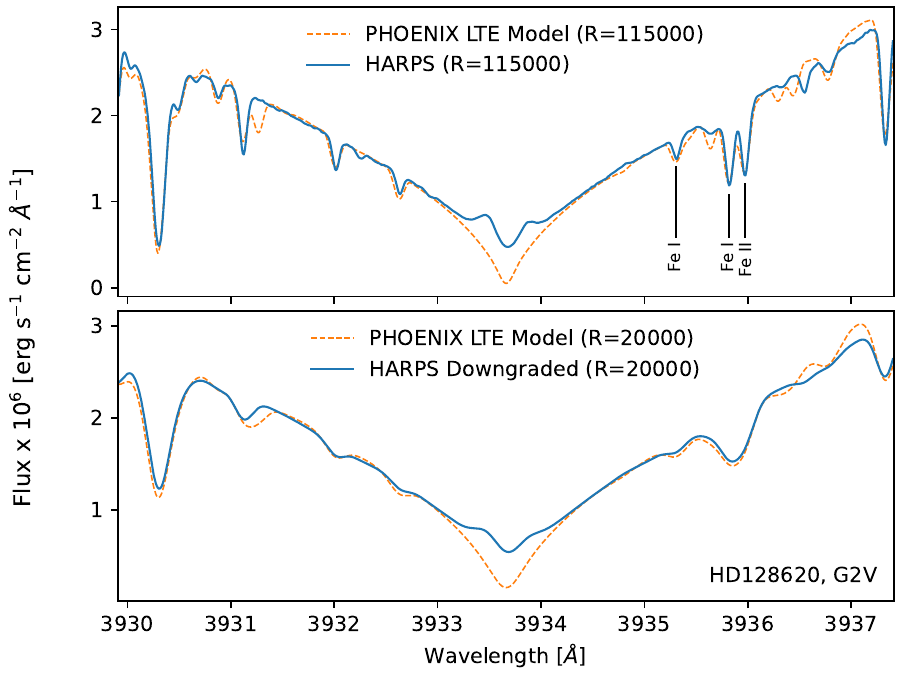}    
    \caption{Comparison of iron emission lines \ion{Fe}{i} and \ion{Fe}{ii} in the wings of \ion{Ca}{ii} observed in HARPS ($R \approx 115,000$) and TIGRE-HEROS ($R \approx 20,000$) spectra. The orange dashed line corresponds to the \textsc{phoenix} photospheric LTE model for each star.}
    \label{fig:HARPS_vs_TIGRE_Fe_and_Ca_II}
\end{figure*}

In Fig. \ref{fig:Synthetic_vs_Observed_Fe_and_CaII}, we revisit that find using a new and homogeneous set of spectra obtained with the Telescopio Internacional de Guanajuato Robótico-Espectroscópico (TIGRE-HEROS) \citep{Schmitt2014, Gonzalez2022}. However, given the modest resolution of the TIGRE-HEROS spectrograph ($R \approx 20,000$), the emission feature measured at 3935.887 \AA{} is in fact a blend of two well–identified transitions of neutral and singly–ionized iron. Using the NIST Atomic Spectra Database \citep{NIST2024}, we identified \ion{Fe}{i} at $\lambda = 3935.812$ \AA{} in air (lower-level: $3d^7$($^4P$)$4s$ b $3P_2$; upper-level: $3d^6$($^3G$)$4s4p$($^3P$°) v $5F$°$_2$) and \ion{Fe}{ii} at $\lambda = 3935.942$ \AA{} in air (lower-level: $3d^6$($^1F$)$4s$ c $^2F_\text{7/2}$; upper-level: $3d^6$($^3G$)$4p$ x $^2G$°$_\text{9/2}$). The Fig. \ref{fig:HARPS_vs_TIGRE_Fe_and_Ca_II} shows both iron lines and a comparison between high-resolution HARPS spectra ($R \approx 115,000$, \citealt{Mayor2003}), TIGRE-HEROS spectra, and synthetic \textsc{phoenix} photospheric LTE models \citep{Hauschildt1999}. In addition, the partner \ion{Fe}{i} line (same lower and upper-level) at 3935.306 \AA{} was also identified. These examples seem to show the transformation of the iron wing-lines from a regular photospheric absorption profile to emission, in proportion to an apparent increasing emission of \ion{Ca}{ii}.

The physical mechanisms that promote the formation of these wing-emission lines in different types of stars has long been debated. Classical radiative–transfer studies \citep{Lites1974, Cram1980, Watanabe1986} showed that \ion{Fe}{ii} wing-emission lines do not require radiative coupling with \ion{Ca}{ii}, but can instead arise from internal iron excitation and from the structure of the low chromosphere. In this work we build on that foundation by examining how the \ion{Fe}{i}+\ion{Fe}{ii} blend correlates with stellar parameters across a large and homogeneous sample. 

The high quality collection of modest high-resolution spectra of F, G, and K giants obtained by TIGRE-HEROS are met with matching \textsc{phoenix} photospheric stellar models to generate both surface fluxes in physical units as well as rigorously determined physical parameters for our stellar sample. In addition to low-noise observational data, this study enabled by methods to determine stellar physical parameters \citep{Rosas-Portilla2022} and chromospheric surface fluxes in absolute terms \citep{Rosas-Portilla2024}. Together, this approach allows us to confirm and quantify that the \ion{Fe}{i}+\ion{Fe}{ii} emission blend in the red wing of \ion{Ca}{ii} K increases proportionally with the \ion{Ca}{ii} emission. For similar stellar parameters, stronger \ion{Ca}{ii} K line emission leads to stronger \ion{Fe}{i}+\ion{Fe}{ii} line emission. This wing-emission blend shows a broad empirical dependence on stellar parameters, including gravity, across different evolutionary stages, similar to \ion{Ca}{ii} K emission. This offers a new empirical approach to understanding how chromospheric structure and iron-line formation vary across the Hertzsprung–Russell diagram (HRD).

\section{Observational data and sample}

\subsection{Forming a representative stellar sample}

We selected 21 well-observed G and K giants spanning luminosity classes (LC) I-III and covering a representative range of effective temperatures (3853–5332 K), surface gravities (0.97–3.41 dex), and solar-like metallicities (see Table \ref{table:stellar_parameters} below). The stellar activity of our sample is mostly low to moderate (see the chromospheric \ion{Ca}{ii} K emission line profiles in the Data Availability section). The rotational and turbulent velocities including their variation throughout the sample remain within the spectral resolution of TIGRE-HEROS, which is equivalent to 7.5 km s$^{-1}$, except for the most active giants whose rotation velocities exceed 8 km s$^{-1}$ \citep{Auriere2015}.

\begin{table*}
	\centering
	\caption{Stellar parameters and chromospheric emission-line flux measurements of \ion{Ca}{ii} K and \ion{Fe}{i} + \ion{Fe}{ii} blend.}
	\label{table:stellar_parameters}
		\begin{threeparttable}
			\begin{tabular}{lllcccccc}
				\hline
Star & Name & Type$^a$ & $T_\text{eff}$$^b$ & $\log g$$^b$ & $\text{[Fe/H]}$$^b$ & $F_\text{Ca II}$$^c$ & $F_\text{Fe I+Fe II}$$^c$ \\
				& & & [K] & [dex] & [dex] & $\times 10^{5}$ [erg s$^{-1}$ cm$^{-2}$] & $\times 10^{5}$ [erg s$^{-1}$ cm$^{-2}$] \\ \hline
HD	8512	&	$\theta$ Cet	&	K0IIIb	&	4765	&	2.64	&	-0.15	&	0.769	&	0.157	\\
HD	18925	&	$\gamma$ Per	&	G9III	&	5083	&	2.29	&	-0.20	&	1.460	&	0.289	\\
HD	27371	&	$\gamma$ Tau	&	G9.5IIIab	&	4998	&	2.72	&	0.07	&	2.170	&	0.23	\\
HD	27697	&	$\delta$ Tau	&	G9.5III	&	4986	&	2.72	&	0.07	&	1.260	&	0.187	\\
HD	28305	&	$\epsilon$ Tau	&	G9.5III	&	4939	&	2.69	&	0.17	&	0.951	&	0.172	\\
HD	28307	&	77 Tau	&	G9III	&	5000	&	2.88	&	0.11	&	1.840	&	0.232	\\
HD	29139	&	$\alpha$ Tau	&	K5+III	&	3853	&	1.27	&	-0.24	&	0.163	&	0.0098	\\
HD	31398	&	$\iota$ Aur	&	K3II-III	&	4050	&	1.13	&	-0.13	&	0.252	&	0.0155	\\
HD	32068	&	$\zeta$ Aur	&	K5II	&	3942	&	0.97	&	-0.14	&	0.215	&	0.0077	\\
HD	48329	&	$\epsilon$ Gem	&	G8Ib	&	4553	&	1.11	&	-0.35	&	1.680	&	0.0917	\\
HD	71369	&	$o$ Uma	&	G5III	&	5235	&	2.66	&	-0.09	&	1.900	&	0.397	\\
HD	81797	&	$\alpha$ Hya	&	K3IIIa	&	4117	&	1.41	&	-0.11	&	0.265	&	0.0177	\\
HD	82210	&	24 Uma	&	G5III-IV	&	5332	&	3.41	&	-0.30	&	13.500	&	0.473	\\
HD	96833	&	$\psi$ Uma	&	K1III	&	4647	&	2.28	&	-0.11	&	0.671	&	0.109	\\
HD	104979	&	$o$ Vir	&	G8III	&	4986	&	2.75	&	-0.32	&	1.320	&	0.321	\\
HD	109379	&	$\beta$ Crv	&	G5II	&	5226	&	2.56	&	0.03	&	1.770	&	0.409	\\
HD	115659	&	$\gamma$ Hya	&	G8IIIa	&	5150	&	2.75	&	0.03	&	1.930	&	0.361	\\
HD	124897	&	$\alpha$ Boo	&	K1,5III	&	4355	&	1.70	&	-0.57	&	0.514	&	0.0585	\\
HD	148387	&	$\eta$ Dra	&	G8-IIIab	&	5083	&	2.86	&	-0.08	&	1.370	&	0.299	\\
HD	186791	&	$\gamma$ Aql	&	K3II	&	4098	&	1.06	&	-0.13	&	0.311	&	0.0199	\\
HD	205435	&	$\rho$ Cyg	&	G8III	&	5153	&	3.05	&	-0.14	&	5.170	&	0.316	\\
				\hline
			\end{tabular}
			\begin{tablenotes}
				\footnotesize
                    \item[a] Spectral type from SIMBAD database.
				\item[b] According to our tests, the uncertainty in effective temperature is $\pm 70$ K, for gravity is $\pm 0.09$ dex and for metallicity is $\pm 0.08$ dex.
                    \item[c] The uncertainty is approximately 10 percent of the emission line flux.
			\end{tablenotes}
		\end{threeparttable}
\end{table*}

\subsection{Homogeneous quality set of stellar spectra}

All observations were made with the 1.2 m robotic telescope TIGRE (located near Guanajuato, central Mexico), which is equipped with the HEROS echelle spectrograph ($R \approx 20,000$ over a wide wavelength range of 3800 to 8800 \AA{}). These spectra were collected with long exposure times (typically $\sim$ 600 - 800 s) to produce a signal-to-noise (S/N) of $\sim$ 50-70 in the near-UV \ion{Ca}{ii} H\&K line region.

In the selection of such suitable spectra, a visual quality control was performed to remove any non-uniform results of our automatized echelle spectra reduction, in case of the results of a particular night were unsatisfactory due to poor transparency or large variability of its photometric quality, which can affect the normalization to a pseudo-continuum. To further improve quality and robustness of the results, we added two or three individual spectra for each star before measuring the \ion{Ca}{ii} K and \ion{Fe}{i}+\ion{Fe}{ii} emission-line fluxes.

\subsection{Precise and consistent parameter determination}

In \cite{Rosas-Portilla2022, Rosas-Portilla2024}, we performed a precise determination of the physical parameters of our sample using a combination of TIGRE-HEROS spectra, \textit{Gaia} DR3 parallaxes \citep{GaiaCollaboration2016, 
GaiaCollaboration2020, GaiaCollaboration2023}, and available public photometry. According to our tests, typical uncertainties are: $T_\text{eff} = \pm 70$ K, $\log g = \pm 0.09$ dex and $\text{[Fe/H]} = \pm 0.08$ dex. For more details of the physical parameter determination and the carefully developed synthesizing method, we refer the interested reader to our previous publications. Table \ref{table:stellar_parameters} summarizes the main physical parameters for the stellar sample stars presented here.

\section{Precision measurements of the chromospheric Fe I+Fe II blend and Ca II K emission-line fluxes}

Synthetic photospheric models using well-determined stellar parameters yield excellent scaling of the surface fluxes in physical units and define the photospheric contribution in the core of a line and thus the zero point for any chromospheric flux measurements. Over three decades of continuous development, \textsc{phoenix} \citep{Hauschildt1999} is a general-purpose state-of-the-art model atmosphere code designed to calculate synthetic atmospheres and spectra of stars and stellar-like objects across the Hertzsprung-Russell diagram. In \cite{Rosas-Portilla2024}, we used \textsc{phoenix} to create a synthetic spectral library of 32 stars, including the stellar sample studied here. In addition, we demonstrated that the differences between the \textsc{phoenix} synthetic models when considering local thermodynamic equilibrium (LTE) and non-local thermodynamic equilibrium (NLTE) are minimal in the wavelength range of interest to this work, i.e. the wings of \ion{Ca}{ii} K line.

We developed a script for \textsc{iSpec} \citep{Blanco-Cuaresma2014, Blanco-Cuaresma2019} written in \textsc{python} 3 to automatically rescale each TIGRE-HEROS spectrum by matching the spectral pseudo-continuum level to the \textsc{phoenix} synthetic model in suitable wavelength regions close to \ion{Ca}{ii} K-line. Subsequently, our script determines an optimal scaling factor by applying the least squares method between the observed and the synthetic spectrum. We included a representation by cubic splines to reduce variations caused by electronic noise. Further details of this method are explained in detail in \cite{Rosas-Portilla2024}. This ensures a consistent absolute flux calibration and defines a clean photospheric baseline for measuring chromospheric emission. 

Thereafter, the emission line fluxes of \ion{Fe}{i}+\ion{Fe}{ii} blend and \ion{Ca}{ii} K can be directly measured in terms of absolute physical flux, i.e. in units of ergs per second per square centimeter. This process basically consists of measuring the area between the chromospheric emission of the stellar spectrum observed with TIGRE-HEROS, and the photospheric absorption of the \textsc{phoenix} synthetic model (see Fig. \ref{fig:Synthetic_vs_Observed_Fe_and_CaII}). However, we note that the \ion{Ca}{ii} K emission-line flux reported in \cite{Rosas-Portilla2024} were computed using an earlier version of our rescaling procedure, which slightly overestimated the absolute calibration. Reprocessing those spectra with the updated pipeline yields systematically lower fluxes, but in very tight proportionality to the original values, following $\log F_{\mathrm{new}} = 0.96\,\log F_{\mathrm{old}} - 1.16$. This linear relation indicates that the revised calibration affects only the absolute zero-point of the \ion{Ca}{ii} K scale, without modifying any relative trends among stars. We adopt the updated flux scale throughout this work. Table \ref{table:stellar_parameters} summarizes the chromospheric emission-line flux for the \ion{Fe}{i}+\ion{Fe}{ii} blend and \ion{Ca}{ii} K.

\section{Emission-line flux relations and their stellar parameter dependence}

The results listed in Table \ref{table:stellar_parameters} allow us to directly compare the \ion{Fe}{i}+\ion{Fe}{ii} and \ion{Ca}{ii} K emission fluxes for stars with similar physical parameters but different chromospheric emission degrees. The stars HD71369 ($o$ UMa, G5III) and HD109379 ($\beta$ Crv, G5IIB) are an example case for which both the \ion{Ca}{ii} K and \ion{Fe}{i}+\ion{Fe}{ii} emission-line flux ratios are in proportion by a factor of 0.93 and 1.03, respectively. Considering the similarity between their effective temperature and surface gravity, these differences in emission are down to different stellar activity levels.

\begin{figure}
    \includegraphics[width=0.99\columnwidth]{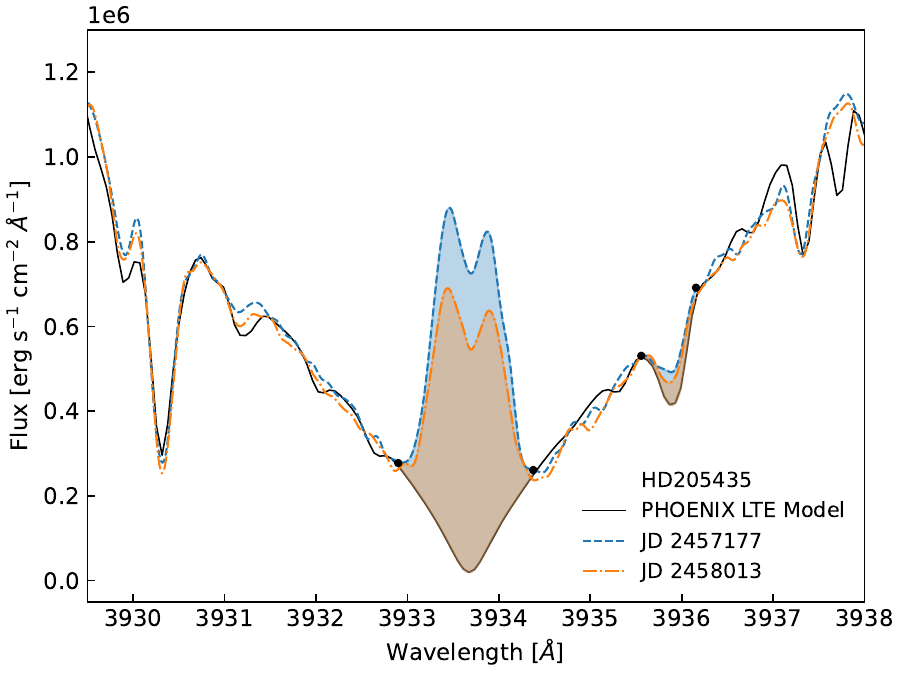}
    \caption{Variability of the emission-line of \ion{Fe}{i} + \ion{Fe}{ii} blend in the \ion{Ca}{ii} K wing observed at two different epochs for the star HD205435 ($\rho$ Cyg, G8III) using TIGRE-HEROS spectra. The points represent the wavelength range in which \ion{Ca}{ii} K and iron blend fluxes were measured. The fluxes in erg s$^{-1}$ cm$^{-2}$ are: (JD 2457177) $F_\text{Ca II} = 6.576 \times 10^5$ and $F_\text{Fe I + Fe II} = 2.799 \times 10^4$; and (JD 2458013) $F_\text{Ca II} = 5.005 \times 10^5$ and $F_\text{Fe I + Fe II} = 2.065 \times 10^4$.}
    \label{fig:FeI_and_CaII_Emission_Variability_HD205435}
\end{figure}

Fig. \ref{fig:FeI_and_CaII_Emission_Variability_HD205435} shows another example of the variability of the \ion{Fe}{i}+\ion{Fe}{ii} blend in proportion to the \ion{Ca}{ii} K emission. Using the TIGRE-HEROS spectra of the star HD205435 ($\rho$ Cyg, G8III) obtained at two different epochs (JD 2457177 and JD 2458013), it is observed that the \ion{Ca}{ii} K and \ion{Fe}{i}+\ion{Fe}{ii} emissions are proportional to each other at a ratio of 0.761 and 0.737, respectively, between the epochs of highest and lowest emission.

\begin{table*}
\centering
\caption{Dependence of the $\log F_\text{FeI + FeII}$ blend emission flux with surface gravity, effective temperature and metallicity.}
\label{table:gravity_and_temperature_dependence_of_Fe_flux}
\begin{tabular}{ccccccc} \hline
    $\alpha$ & $\beta$ & $\gamma$ & $C$ & $\chi^2_\nu$ & $R^2$ \\ \hline
    0 & 13.0 (0.3) & -0.28 (0.08) & -43.6 (1.2) & 17.907 & 0.989 \\
    0.71 (0.08) & 0 & -0.4 (0.3) & 2.43 (0.19) & 259.014 & 0.829 \\
    -0.03 (0.06) & 13.1 (1.0) & 0 & -44 (4) & 27.403 & 0.982 \\
    0.00 (0.05) & 13.0 (0.8) & -0.28 (0.09) & -43.6 (2.9) & 16.912 & 0.989 \\ \hline
\end{tabular}
\end{table*}

The approximate proportionality between the \ion{Fe}{i} + \ion{Fe}{ii} wing emission and the \ion{Ca}{ii} K flux is consistent with previous radiative–transfer studies \citep{Lites1974, Cram1980}, which showed that metallic emission lines in the \ion{Ca}{ii} H\&K wings respond primarily to the thermal and density structure of the lower chromosphere rather than to any direct radiative pumping by \ion{Ca}{ii} H\&K. Under this interpretation, both diagnostics trace the same underlying chromospheric conditions, explaining the tight empirical relation observed among stars with similar stellar parameters, and also within individual stars across different epochs. Therefore, it is pertinent to investigate whether the \ion{Fe}{i} + \ion{Fe}{ii} blend follows comparable empirical dependencies on the stellar parameters.

In \citet{Rosas-Portilla2024}, it was shown that the \ion{Ca}{ii} K emission-line flux follows a power-law relation with surface gravity and effective temperature, reflecting how chromospheric heating, density stratification, and ionization balance scale across different evolutionary stages. If the \ion{Fe}{i} + \ion{Fe}{ii} blend indeed acts as a chromospheric indicator in a similar formation region, its flux should exhibit analogous power-law trends. Testing this expectation provides a direct way to assess whether both diagnostics respond coherently to changes in the lower chromosphere. To explore this possibility, we adopt a similar functional form introduced by \citet{Rosas-Portilla2024} for \ion{Ca}{ii} K, fitting the \ion{Fe}{i} + \ion{Fe}{ii} blend emission flux with a simple power-law in the following logarithmic form:

\begin{equation}
    \log F_\text{FeI + FeII} = \alpha \log\,g + \beta \log T_\text{eff} + \gamma \text{[Fe/H]} + C,
    \label{eq:log_F_vs_logg_and_log_teff}
\end{equation}

\noindent where $\alpha$, $\beta$, $\gamma$, and $C$ are fitting constants. We introduce an additional term that depends on metallicity ([Fe/H]), since our analysis is about an iron-based feature and the sample spans a nonnegligible range in [Fe/H]. Table \ref{table:gravity_and_temperature_dependence_of_Fe_flux} shows the results for these values when we apply Eq. \ref{eq:log_F_vs_logg_and_log_teff} considering the special cases when $\alpha = 0$ (i.e., a "pure" dependence on effective temperature and metallicity), $\beta = 0$ (i.e., a "pure" dependence on gravity and metallicity), and $\gamma = 0$ (i.e., a "pure" dependence on gravity and effective temperature).

Based on the results of the reduced chi-square ($\chi^2_\nu$) and coefficient of determination ($R^2$) for a best-matched empirical relation performed by a least-squares fit, we observed that the effective temperature exponent is similar to that found for the \ion{Ca}{ii} K emission-flux in \citet{Rosas-Portilla2024} (see Table 2), whereas the surface gravity exponent is statistically zero when the entire stellar sample is considered. However, as we show below, the dependence on gravity is different for different regimes of this parameter. In addition, we find evidence of an additional and not negligible dependence on metallicity. This suggests that, within the parameter range of our sample, the \ion{Fe}{i} + \ion{Fe}{ii} blend may reflect underlying chromospheric thermal conditions in an indirect way. In contrast, the \ion{Ca}{ii} K line, being a strong resonance transition, retains a clearer dependence on chromospheric column density and gravity, as reflected in its non-zero $\alpha$-exponent (see \citealt{Rosas-Portilla2024}). It is important to mention that reduced chi-square values exceeding the unity may indicate that uncertainties in the flux measurements are somewhat underestimated, or that the intrinsic temporal variability of the chromospheric emission, not considered in our model, contributes to the observed dispersion.

\section{The Fe I + Fe II emission-line flux ratio as a differential chromospheric diagnostic}

The \ion{Ca}{ii} K emission-line and the \ion{Fe}{i} + \ion{Fe}{ii} wing emission probe similar regions of the lower chromosphere but with notably different sensitivities: \ion{Ca}{ii} K responds to changes in column density and chromospheric extension, whereas the \ion{Fe}{i} + \ion{Fe}{ii} seems to be more related to the temperature structure. Comparing their emission-line fluxes therefore provides a differential diagnostic capable of tracing how thermal and structural properties vary across our sample. In particular, the emission flux ratio (\ion{Fe}{i} + \ion{Fe}{ii})/\ion{Ca}{ii} can reveal whether the iron-blend increases or decreases more rapidly than the \ion{Ca}{ii} K flux as a function of stellar parameters, and consequently offers insights into the relative formation heights of both indicators.

\begin{table*}
\centering
\caption{Dependence by groups of the $\log F_\text{FeI + FeII}$ blend emission flux with surface gravity, effective temperature and metallicity.}
\label{table:gravity_and_temperature_dependence_of_Fe_flux_by_groups}
\begin{tabular}{ccccccccc} \hline
    Group & Condition & $\alpha$ & $\beta$ & $\gamma$ & $C$ & $\chi^2_\nu$ & $R^2$ \\ \hline
    \multirow{4}{*}{G1 (12 stars)} & \multirow{4}{*}{$\log g > 2.5$} & 0 & 9.9 (1.2) & -0.26 (0.09) & -32 (4) & 9.934 & 0.918 \\
    & & 0.14 (0.22) & 0 & -0.39 (0.31) & 4.1 (0.6) & 85.247 & 0.295 \\
    & & -0.03 (0.10) & 11.1 (1.7) & 0 & -37 (6) & 18.061 & 0.851 \\
    & & -0.15 (0.07) & 10.9 (1.1) & -0.34 (0.09) & -35 (4) & 7.053 & 0.948 \\ \hline
    \multirow{4}{*}{G2 (9 stars)} & \multirow{4}{*}{$\log g \leq 2.5$} & 0 & 13.1 (0.6) & -0.44 (0.16) & -44.3 (2.3) & 20.925 & 0.989 \\
    & & 0.80 (0.24) & 0 & -0.9 (0.8) & 2.2 (0.4) & 465.928 & 0.698 \\
    & & 0.02 (0.12) & 13.4 (1.5) & 0 & -45 (5) & 43.113 & 0.972 \\
    & & 0.05 (0.08) & 12.6 (1.1) & -0.45 (0.17) & -42 (4) & 18.685 & 0.988 \\
    \hline
\end{tabular}
\end{table*}

We relate the above emission flux ratio to the three fundamental stellar parameters potentially of impact here: (i) the surface gravity, $\log g$; ii) the effective temperature, $T_\text{eff}$; and metallicity, ([Fe/H]). In the first case, the gravity varies over four orders of magnitude, and spreads chromospheric column densities (growing approximately with $g^{-0.5}$) \citep{Ayres1975} by at least two orders of magnitude. This has an impact on the respective line formation and optical depths, on ionization, and consequently, on the available numbers of neutral iron scatterers. On the other hand, for the second and third case, $T_\text{eff}$ and [Fe/H] are important factors in ionization in the relevant chromospheric line formation region. However, in sharp contrast to gravity, the ranges of effective temperature and metallicity are quite narrow.

\begin{figure}
	\includegraphics[width=0.99\columnwidth]{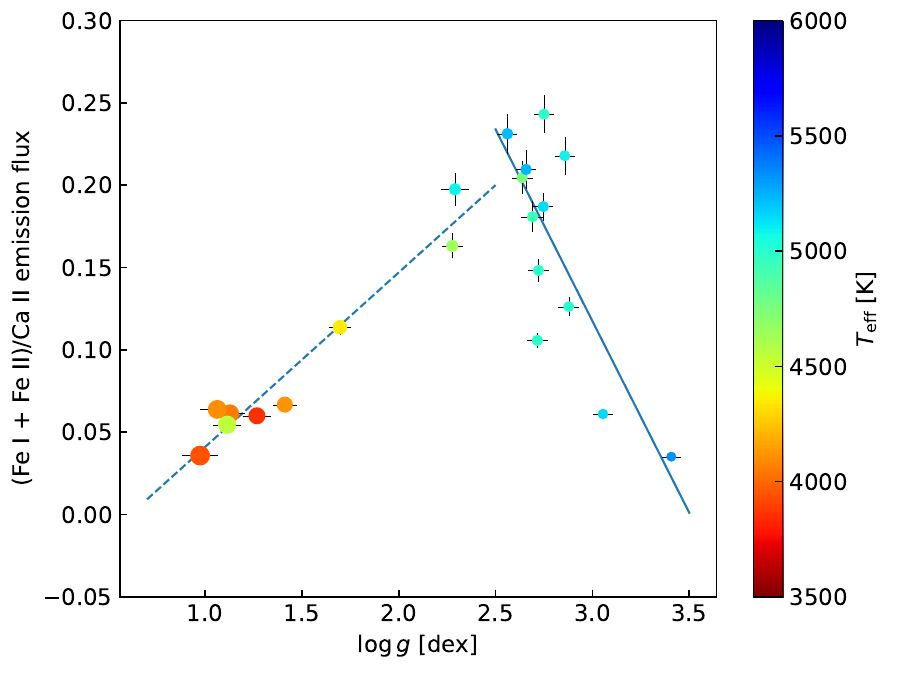}
	\caption{Dependence of the  (\ion{Fe}{i} + \ion{Fe}{ii})/\ion{Ca}{ii} K emission line flux ratio on surface gravity. The blue solid and dashed lines correspond to a linear fit for stars with $\log g > 2.5$ dex (Group 1) and $\log g \leq 2.5$ dex (Group 2) respectively. The color code represents the effective temperature ($T_\text{eff}$), and the point size represents the surface gravity ($\log g$), with larger points for lower gravities.}
	\label{fig:Area_Fe_Area_K_vs_log_g}
\end{figure}

Our Fig. \ref{fig:Area_Fe_Area_K_vs_log_g} illustrates the dependence of the (\ion{Fe}{i} + \ion{Fe}{ii})/\ion{Ca}{ii} emission line flux ratio as a function of surface gravity. In this relation, two different trends are observed for stars with $\log g > 2.5$ dex and $\log g \leq 2.5$ dex. The inflection point around $\log g \approx 2.5$ dex may illustrate a change in the balance of the physical conditions that regulate both diagnostics. To further investigate the origin of this apparent change, we explore whether the dependence of the \ion{Fe}{i} + \ion{Fe}{ii} wing emission on stellar parameters differs between these two regimes. Therefore, we divided the sample into two groups, defined by Group 1: $\log g > 2.5$ dex; and Group 2: $\log g \leq 2.5$ dex. Table \ref{table:gravity_and_temperature_dependence_of_Fe_flux_by_groups} shows the results when we apply the same logarithmic function introduced in Eq. \ref{eq:log_F_vs_logg_and_log_teff}. We performed separate fits for each group to assess whether the influence of these parameters changes across the transition. This also allows us to test whether the observed inflection reflects a genuine change in the underlying physical conditions.

From the results of the reduced chi-square ($\chi^2_\nu$) and coefficient of determination ($R^2$) for a best-matched empirical relation performed by a least-squares fit, we note that this behavior seems to hint at a structural transition in the chromospheres of stars from LC I to III. For lower gravities, \ion{Ca}{ii} K is expected to become increasingly sensitive to the larger chromospheric extension \citep{Berio2011, Diamant2023} and density scale height, while the \ion{Fe}{i} + \ion{Fe}{ii} blend --formed in a narrower layer of the lower chromosphere-- continues to track the local source function. The upward deviation of the ratio at $\log g \approx 2.5$ may therefore signal a shift between a regime where both diagnostics reflect similar physical conditions and one where they probe distinct aspects of the chromospheric structure.

\section{Discussion and conclusions}

The precise determination of the stellar parameters for our sample in combination with well-calculated \textsc{phoenix} synthetic spectra allows us to measure the absolute chromospheric emission-line fluxes of the \ion{Ca}{ii} K and the \ion{Fe}{i} + \ion{Fe}{ii} blend in a set of homogeneous and high-quality spectra. Reliable absolute fluxes are essential to revisit the long-known phenomenon of wing emission and to analyze its dependence on fundamental stellar parameters.

In this work, we first confirm the stars with similar physical parameters and as such with similar chromospheric conditions, but different chromospheric activity levels as traced by the \ion{Ca}{ii} K emission-line flux, show a clear proportionality with the \ion{Fe}{i} + \ion{Fe}{ii} blend wing emission. This behavior is consistent for different luminosity classes, and within individual stars at different epochs, showing that the iron-blend seems to respond to changes in chromospheric conditions in parallel with \ion{Ca}{ii} K.

Second, our empirical power-law analysis shows that the \ion{Fe}{i} + \ion{Fe}{ii} blend flux is controlled by effective temperature, with a temperature exponent comparable to that found for \ion{Ca}{ii} K in \citet{Rosas-Portilla2024}. In contrast, the gravity exponent is statistically zero when the entire stellar sample is considered; however, it differs between stars with surface gravities lower and higher than 2.5 dex. This could indicate, at least within the range of our sample (LC I–III), that the iron blend seems to behave as a proxy for chromospheric thermal conditions while showing little sensitivity to global chromospheric extension. It is important to note that our analysis is based on stellar effective temperature and does not directly probe the chromospheric temperature structure. This interpretation is consistent with constraints from chromospheric-eclipse studies of $\zeta$ Aur (HD 32068) and 31 Cyg (HD 192577) (González-Enríquez et al., in preparation), which show that many metallic lines used in curves of growth, including \ion{Fe}{i} and \ion{Fe}{ii} transitions, are sensitive to conditions in the narrow region of the lower chromosphere, with well-defined gradients in temperature and density. This localized formation region is expected to respond strongly to local thermal conditions and explains the interpretation that iron-line diagnostics primarily trace the local thermal structure of this layer. A more direct assessment of the chromospheric temperature sensitivity would require detailed radiative-transfer modeling, which is beyond the scope of the present work.

Third, the (\ion{Fe}{i} + \ion{Fe}{ii}) / \ion{Ca}{ii} emission-flux ratio as a function of surface gravity reveals a chromospheric structure with opposite behavior on either side of $\log g \approx 2.5$ suggesting a change in the relative response of both diagnostics. We re-examine the dependence of the \ion{Fe}{i} + \ion{Fe}{ii} blend on the stellar parameters but using the boundary of $\log g \approx 2.5$ dex to separate into two groups. For Group 1: $\log g > 2.5$ dex, we see a dependence of the iron blend on surface gravity. On the other hand, it is evident the lack of dependence on surface gravity for Group 2: $\log g \leq 2.5$ dex. It implies that at lower gravities, where chromospheres are more extended, \ion{Ca}{ii} K becomes increasingly sensitive to the growing column density, while the iron-blend --formed in a narrower layer of the lower chromosphere-- continues to trace mainly the local source function. At higher gravities, the ratio declines again, indicating that \ion{Ca}{ii} K weakens less than the iron-blend as the chromosphere becomes more compact. Although the present stellar sample is limited, this pattern points to the Fe-blend/\ion{Ca}{ii} ratio as a sensitive differential diagnostic of chromospheric structure across the red giant branch.

The peak and reversal of the (\ion{Fe}{i} + \ion{Fe}{ii})/\ion{Ca}{ii} flux ratio around giants ($\log g \approx 2.5$) falls in the HRD with the coronae-wind dividing lines \citep{Linsky1979, Mullan1982}. By our current understanding, these depict a change, driven by the growing chromospheric extent towards more luminous giants with lower gravity \citep{Diamant2023}. Where cool winds, emerge from extended chromospheres, coronal emission becomes buried \citep{Ayres2003}. At the top of the chromosphere, plasma becomes energetically unable to reach beyond the critical region where radiative processes alone are insufficient to maintain thermal equilibrium \citep{Athay1976} and consequently fails to transition to coronal temperatures \citep{Schroder2018}. The average orientation of the coronal magnetic field loops in the scattering layers plays a role --neutral and ionized iron being a magnetically sensitive atom--. In such a case, the K-line wing \ion{Fe}{i} + \ion{Fe}{ii} blend emission and related phenomena can provide insight into the magnetic structuring of cool giant chromospheres, testable with stellar surface imaging.

\section*{Acknowledgements}

This study used the services of the Strasbourg astronomical data center, and data from the European Space Agency (ESA) mission \textit{Gaia} (\url{https://www.cosmos.esa.int/gaia}), processed by the \textit{Gaia} Data Processing and Analysis Consortium (DPAC, \url{https://www.cosmos.esa.int/web/gaia/dpac/consortium}). This work benefited from financial support of the bilateral project CONACyT-DFG No. 278156 for the TIGRE collaboration with the University of Hamburg, and from general support of our home institutions. The authors thank the anonymous referee for his/her kind suggestions to improve the results shown here. F.R.P. thanks the DGAPA of UNAM (Mexico) for the postdoctoral fellowship that allowed the development of this work. FRP thanks LML, who accompanied an important part of this journey and whose presence will endure beyond time and distance.

\section*{Data Availability}

The rescaled profiles of the emission-lines of \ion{Ca}{ii} K and \ion{Fe}{i} + \ion{Fe}{ii} blend are available at: \url{https://bit.ly/4fXYYiz}.


\bibliography{bibliography}

\end{document}